\documentclass[pre,twocolumn,superscriptaddress,nofootinbib,floatfix]{revtex4-1}
\usepackage{amsmath}
\usepackage{graphicx}
\usepackage{color}
\usepackage{longtable}

\begin{document}

\title{Edge of Chaos and Genesis of Turbulence}

\author{Abraham C.-L. Chian}
\email{abraham.chian@gmail.com}
\affiliation{National Institute for Space Research (INPE) and World Institute for Space Environment Research (WISER), P. O. Box 515, S\~ao Jos\'e dos Campos--SP 12227--010, Brazil}
\affiliation{Observatoire de Paris, LESIA, CNRS, 92195 Meudon, France}
\affiliation{Institute of Aeronautical Technology (ITA), CTA/ITA/IEFM, S\~ao Jos\'e dos Campos--SP 12228--900, Brazil}

\author{Pablo R. Mu\~noz}
\affiliation{National Institute for Space Research (INPE) and World Institute for Space Environment Research (WISER), P. O. Box 515, S\~ao Jos\'e dos Campos--SP 12227--010, Brazil}
\affiliation{Institute of Aeronautical Technology (ITA), CTA/ITA/IEFM, S\~ao Jos\'e dos Campos--SP 12228--900, Brazil}

\author{Erico L. Rempel}
\affiliation{National Institute for Space Research (INPE) and World Institute for Space Environment Research (WISER), P. O. Box 515, S\~ao Jos\'e dos Campos--SP 12227--010, Brazil}
\affiliation{Institute of Aeronautical Technology (ITA), CTA/ITA/IEFM, S\~ao Jos\'e dos Campos--SP 12228--900, Brazil}

\date{\today}

\begin{abstract}
The edge of chaos is analyzed in a spatially extended system, modeled by the regularized long-wave equation, prior to the transition to permanent spatiotemporal chaos.  In the presence of coexisting attractors, a chaotic saddle is born at the basin boundary due to a smooth-fractal metamorphosis. As a control parameter is varied, the chaotic transient evolves to well-developed transient turbulence via a cascade of fractal-fractal metamorphoses. The edge state responsible for the edge of chaos and the genesis of turbulence is an unstable travelling wave in the laboratory frame, corresponding to a saddle point lying at the basin boundary in the Fourier space.

\vspace{0.6cm}
\emph{This paper has been published in Physical Review E 88, 052910 (2013)}
\end{abstract}

\maketitle

\section{Introduction}

There is a growing interest in edge states at the laminar-turbulent boundary which can improve our understanding of the transition from turbulent to laminar flows in fluids and plasmas, as well as the precursors of turbulence \cite{skufca2006,*kim2008,schneider2007, *vanveen2011, *delozar2012, *cherubini2011}. Recently, an interior crisis was fully characterized in the chaotic dynamics of the Pierce diode, a simple spatially extended system for collisionless bounded plasmas, based on the concept of edge of chaos (EOC) \cite{munoz2012}. EOC is defined as the boundary that divides the phase space in two pseudobasins: a region whose initial conditions display a chaotic transient behavior and another region whose initial conditions converge directly to a laminar attractor. The EOC is the stable manifold of an invariant saddle structure called edge state (ES) \cite{skufca2006,schneider2007,munoz2012} and can be obtained by refined techniques such as the bisection method \cite{skufca2006} that allows one to follow the EOC for longer times. For example, in the EOC of a parallel shear flow the ES, determined by the bisection method, is an unstable periodic orbit for low Reynolds numbers whereas at higher Reynolds numbers it is a chaotic object known as the relative attractor \cite{skufca2006}; the ES associated with the EOC in a periodic window of the Pierce diode, found by the bisection method, is a period-3 unstable periodic orbit arisen from a saddle-node bifurcation \cite{munoz2012}. 

The regularized long-wave equation (RLWE), also known as the Benjamin-Bona-Mahony equation, is of great interest in the study of propagation of long waves in shallow waters such as tsunami driven by an earthquake \cite{toledo2013} and drift waves in a controlled nuclear fusion plasma \cite{he1989,*horton1990}. The RLWE is an improved model of nonlinear small-amplitude long-waves in fluids, first derived by \citet{peregrine1966}, then by \citet{benjamin1972} to remove some mathematical problems associated with the Kortweg-de Vries equation, such as the existence and stability of solutions and other problems related to the dispersion term. Dynamical systems description of the transition from temporally to spatiotemporally chaotic attractors, based on the RLWE, provides a simple model to acquire in-depth insights on the laminar-turbulence transition \cite{he2003,rempel2007,chian2010}. 

In this paper we use the RLWE to study the nonlinear dynamics of a spatially extended system prior to the onset of permanent spatiotemporal chaos. The aims are threefold. First, we establish the link between the concept of EOC at the boundary of laminar-turbulent transition and the concept of chaotic saddle at the basin boundary of coexisting attractors. Second, we show that a chaotic saddle is born in a smooth-fractal metamorphosis which evolves to well-developed transient turbulence via fractal-fractal metamorphoses. Third, we elucidate the role of the edge state at the basin boundary of coexisting attractors and at the boundary of pseudo basins of coexisting chaotic saddle and attractor before the onset of permanent spatiotemporal chaos, and at the boundary of pseudo basins of coexisting chaotic saddles/attractors after the onset of permanent spatiotemporal chaos.

\section{The Model}

The driven-damped regularized long-wave equation in dimensionless units is given by \cite{he2003,chian2010}
\begin{equation}
\partial_tu+c\partial_xu+fu\partial_xu+a\partial_{txx}u=-\nu u-\epsilon\sin(\kappa x-\Omega t)
\label{eq_01}
\end{equation}
where $\epsilon$ is the driver amplitude, $c=1$, $f=-6$, $a=-0.28711$, $\nu=0.1$, $\kappa=1$ and $\Omega=0.65$. We impose periodic boundary conditions $u(x,t) = u(x + 2\pi, t)$ and solve Eq. \eqref{eq_01} numerically using a pseudospectral method by expanding the wave variable $u(x,t)$ in a Fourier series
\begin{equation}
u(x,t)=\sum_{k=-N}^{N}\hat{u}_k(t)\exp(ikx).
\label{eq_02}
\end{equation}
We set the number of modes $N=32$ \cite{rempel2007}. By introducing \eqref{eq_02} into Eq. \eqref{eq_01} we obtain a set of ordinary differential equations for the complex Fourier amplitudes $\hat u_k(t)$,
\begin{multline}
(1-ak^2)\frac{d\hat u_k}{dt}=-ick\,\hat u_k-\nu \hat u_k+\frac{\epsilon}2[(\sin\Omega t+i\cos\Omega t)\delta_{1,k}\\
-f{\cal{F}}(u\partial_x u),
\label{eq_03}
\end{multline}
where the last term on the right-hand side is the Fourier transform of the nonlinear part of Eq. \eqref{eq_01}. The pseudo--spectral method computes this term in the real space using the information from the Fourier space. First, we compute the spatial derivative in the Fourier space $\partial_xu\rightarrow ik\hat{u}_k$ and then both $\hat{u}_k$ and $ik\hat{u}_k$ are Fourier-transformed to the real space, where the multiplication $fu \partial_xu$ is performed. Finally, the result is Fourier-transformed back to the Fourier space. Numerical integration is performed using a fourth--order Runge--Kutta integrator, with a time step $\Delta t=T/500$, where $T=2\pi/\Omega$ is the driver period in Eq. \eqref{eq_01}. Since $u(x,t)$ is a real function, only $k>0$ need to be considered and at each time step, 1/3 of the high $k$ modes are set to zero to avoid aliasing errors. Thus,  the effective number of modes is $N=20$ and the phase space has dimension 40, with the state of the system at time $t$ given by $\mathbf{u}=\{\hat u_1,\dots,\hat u_{20}\}$, where $\hat u_k$ is the $k$-th complex Fourier amplitude.

As noticed by \citet{he2005}, Eq. \eqref{eq_01} has solutions of the form $u(x,t)=\tilde u(x-\Omega t)$, which are travelling waves in the laboratory frame $(x,t)$. This kind of solution is a fixed point for the amplitude-phase description of the Fourier modes $\hat u_k$, when it is transformed to the driver frame of reference $\xi=x-\Omega t$. The amplitude and phase of the $k$-th Fourier mode in this frame are given by
\begin{equation}
|\hat u_k|=\sqrt{[\text{Re }\hat{u}_k]^2+[\text{Im }\hat{u}_k]^2}\quad\text{and}  \quad \theta^D_k=\theta^L_k+k\Omega t,
\label{eq_amp_phase}
\end{equation}
where $\theta^L_k=\tan^{-1}\left({\text{Im }\,\hat{u}_k/\text{Re }\hat{u}_k}\right)$ is the phase in the laboratory frame of reference.

\section{Nonlinear Analysis}

In the absence of driving-dissipation ($\epsilon=\nu=0$) or when driving-dissipation is relatively weak ($\epsilon,\nu< 1$), Eq. \eqref{eq_01} admits a steady wave solution in the form of a solitary traveling wave \cite{he1988}. If we keep all parameters in Eq. \eqref{eq_01} fixed and only vary $\epsilon$, the steady wave solution of Eq. \eqref{eq_01} eventually becomes unstable and undergoes a diversity of bifurcations, giving rise to a wealth of dynamical phenomena.
\subsection{Edge of chaos and edge state}
\begin{figure}[t]
\begin{center}
  \includegraphics[width=1\linewidth]{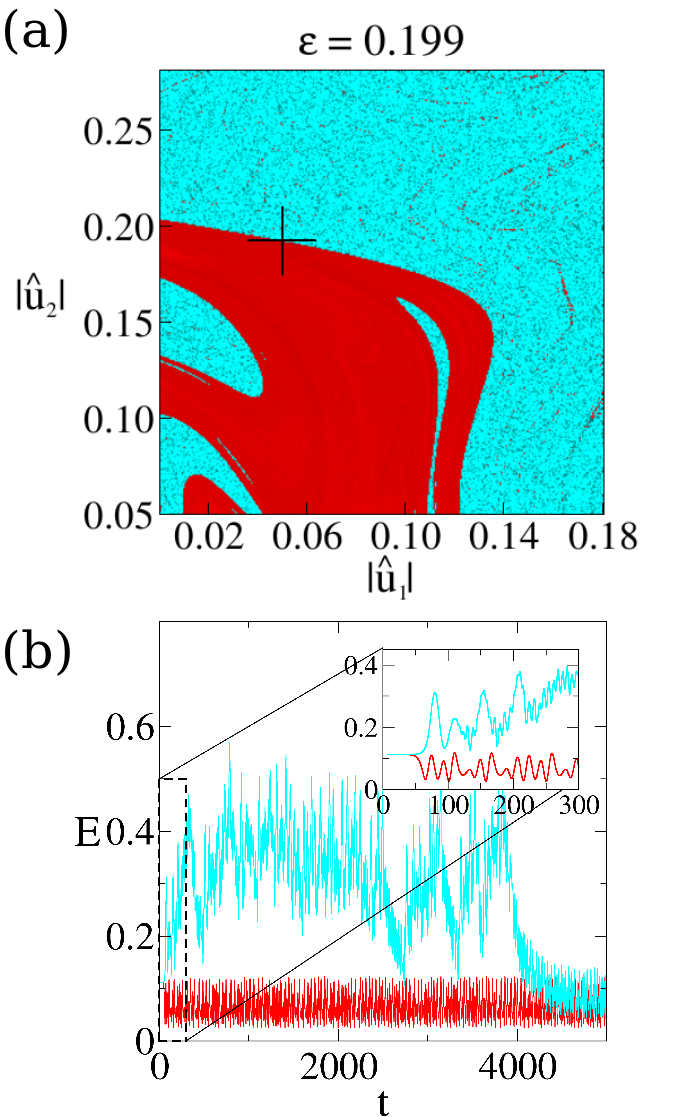}
\end{center}
\caption{(Color online) Edge of chaos is the boundary that separates two regions in the 2D projection of the phase space in (a), at $\epsilon = 0.199$, showing the transient lifetime for turbulent trajectories to converge to a laminar attractor. Edge state is indicated by a black cross whose stable manifold is the edge of chaos. The blue (red) regions indicate long (short) lifetimes that correspond to the initial conditions that do (do not) exhibit transient turbulence before converging to a laminar attractor, as illustrated by the time series of energy $E$ in (b), obtained by the bisection method for two different initial conditions.}
\label{fig_01}
\end{figure}

At $\epsilon = 0.199$, just before the onset of permanent spatiotemporal chaos, the solutions of Eq. \eqref{eq_01} exhibit the characteristics of edge of chaos. A technique to detect the edge of chaos is to compute the lifetime of initial conditions in some region of the phase space \cite{skufca2006, schneider2007}, defined as the time a trajectory takes to converge to the laminar attractor. We construct a two-dimensional projection of the phase space starting from a given initial condition $\mathbf{u}_0$ and varying the amplitude of the first two Fourier modes $|\hat u_1|$ and $|\hat u_2|$ to generate a grid of initial conditions in the driver frame, keeping the other 18 Fourier amplitudes and 20 phases the same as $\mathbf{u}_0$. Figure \ref{fig_01}(a) shows the lifetime landscape in this grid. The base initial condition $\mathbf{u}_0$ is indicated by the black cross in Fig. \ref{fig_01}(a) and the method to find it is explained below. The red regions indicate short lifetimes, and correspond to initial conditions whose trajectories do not show the features of transient turbulence (governed by a spatiotemporally chaotic saddle STCS \cite{rempel2007}) and converge quickly to the laminar attractor (spatially regular and temporally chaotic attractor). On the other hand, the light blue regions correspond to initial conditions whose temporal evolution displays long chaotic transients before converging to the laminar attractor. The stable manifold of STCS is well approximated by the regions of longer lifetime. The edge of chaos is the boundary dividing the two regions of lifetime in Fig. \ref{fig_01}(a).

The cross in Fig. \ref{fig_01}(a) marks the position of the edge state, which lies on the edge of chaos. The edge state is found through the bisection method \cite{skufca2006}. By integrating many different initial conditions it is seen that the trajectories associated with transient turbulence have high level of energy bursts; here, energy is defined by 
\begin{equation}
E =\frac{1}{4\pi}\int_0^{2\pi}[u(x,t)^2-au_x(x,t)^2]\,dx.
\label{eq_04}
\end{equation}
In contrast, the trajectories that converge quickly to the laminar attractor have low level of energy fluctuations. Beginning with two initial conditions $\mathbf{u}_\text{S}$ and $\mathbf{u}_\text{L}$, with short and long lifetimes, respectively, we integrate the condition given by the middle point of the path that connects both conditions, $\mathbf{u}_\text{M}= (\mathbf{u}_\text{S} + \mathbf{u}_\text{L})/2$, until it converges to the laminar attractor. We set the energy level $E_0 = 0.2$ as a threshold to decide to which region of the phase space $\mathbf{u}_{\text{M}}$ belongs. If the maximum energy along the trajectory of $\mathbf{u}_\text{M}$ is lower than $E_0$, $\mathbf{u}_\text{M}$ lies in the laminar pseudo-basin and at the next step we set $\mathbf{u}_\text{L}=\mathbf{u}_\text{M}$. Otherwise, $\mathbf{u}_\text{M}$ belongs to the turbulent pseudo-basin, hence at the next step $\mathbf{u}_S=\mathbf{u}_M$. Repeating this procedure, we find pairs of conditions at both sides of the edge of chaos, arbitrarily close to each other. Figure \ref{fig_01}(b) shows an example of two initial conditions determined by the bisection method, with the distance between them $||\mathbf{u}_\text{S}- \mathbf{u}_\text{L}|| < 10^{-12}$. The red curve is the trajectory of laminar condition $\mathbf{u}_\text{S}$, and the light blue curve is the trajectory of turbulent condition $\mathbf{u}_\text{L}$. As the inset in Fig. \ref{fig_01}(b) shows, both trajectories remain close to each other initially with $E$ remaining almost constant until $t\sim50$. That part of the solutions corresponds to trajectories passing near the stable manifold (EOC) of ES, which is a saddle fixed point in the amplitude-phase description of Fourier space in the driver frame, with constant energy for a given control parameter. For $t\gtrsim 50$, the 2 trajectories separate quickly. The laminar trajectory converges immediately to the laminar attractor, while the turbulent trajectory traverses first the vicinity of a chaotic saddle before converging to the laminar attractor at $t\sim 4000$. By applying systematically the bisection method it is possible to find a long trajectory close to ES. 

\begin{figure}[b]
\begin{center}
\includegraphics[width=1\linewidth]{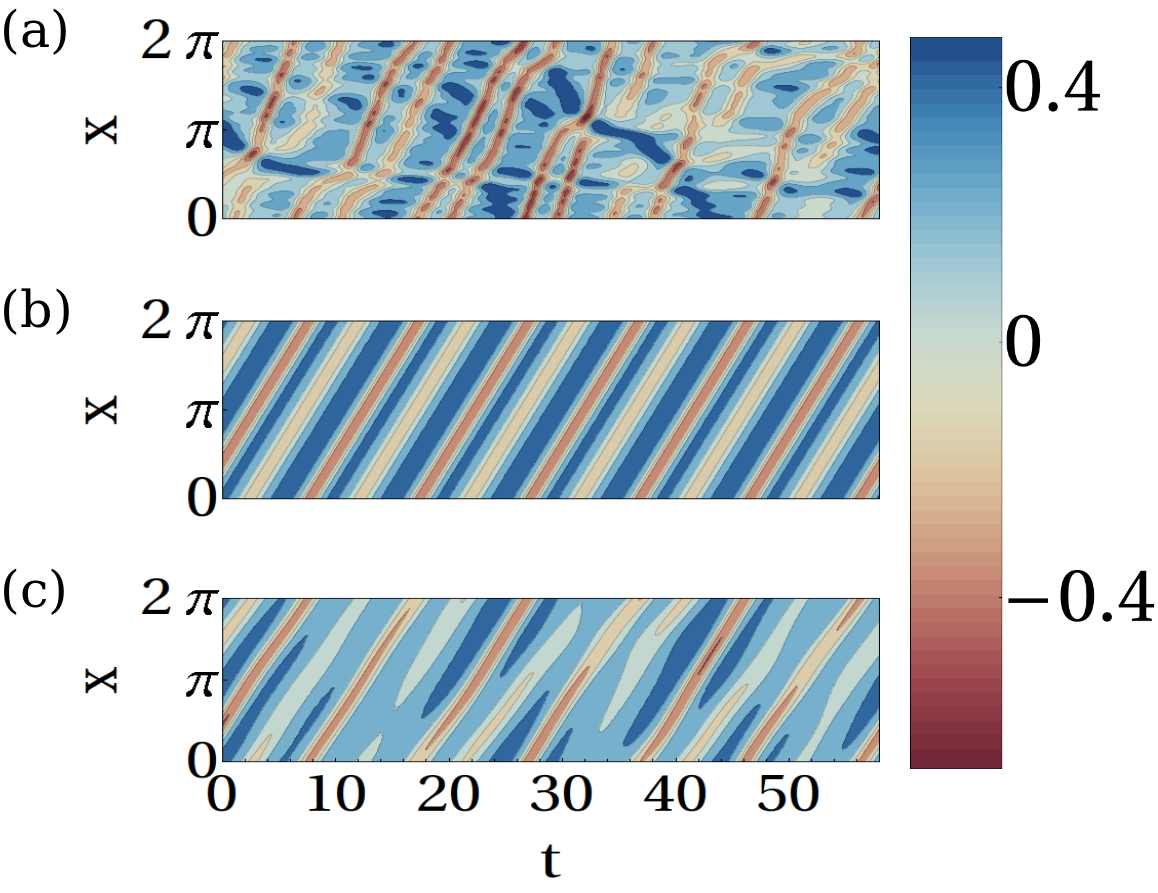} 
\end{center}
\caption{(Color online) Space-time contour plot of $u(x,t)$ in the laboratory frame for three dynamical structures related to the edge of chaos at $\epsilon = 0.199$: (a) transient turbulence, (b) edge state, and (c) laminar attractor.}
\label{fig_02}
\end{figure}

The space-time contour plots in the laboratory frame of three dynamical structures connected to EOC at $\epsilon=0.199$ are shown in Fig. \ref{fig_02}. We characterize the degree of spatiotemporal disorder of each structure by computing the Lyapunov spectrum $\{\lambda_j\}$ and the Kaplan-Yorke dimension \cite{ott1993,chian2010}, defined as
\begin{equation}
D=p+\sum_{j=1}^p\frac{\lambda_j}{\lambda_{p+1}},
\end{equation}
where $p=\max\{ m \,|\, \sum_{j=1}^m\lambda_j\geq 0 \}$. 

We use the stagger-and-step method \cite{sweet2001} to obtain the transient turbulence (STCS) of Fig. \ref{fig_02}(a). Considering that the lifetime of any state in the chaotic saddle STCS is infinite, the stagger-and-step method consists of integrating a piecewise continuous trajectory containing points whose lifetime is greater than some typically large threshold $T_c$. First, we search for an initial condition $\mathbf{u}_0$ at $t = t_0$ with lifetime $T(\mathbf{u}_0)>T_c$ (for instance, some initial condition in the blue region of Fig. \ref{fig_01}(a)) and integrate it until time $t_1 = t_0 + T(\mathbf{u}_0)-T_c$, saving the trajectory as being part of the STCS. We define a new initial condition $\mathbf{u}_1 = \mathbf{u}(t_1)$, with lifetime $T(\mathbf{u}_1)=T_c$, and generate random perturbations $\mathbf{r}$ such that $T(\mathbf{u}_1 +\mathbf{r})>T_c$. \citet{sweet2001} found that the random search is faster when $||\mathbf{r}|| = 10^{-s}$, with $s$ being a uniformly distributed random number between 3 and the machine precision, 15 in our case. The perturbation $\mathbf{r}$ which increases the lifetime of the initial condition $\mathbf{u}_1$ is called a ``stagger'', and the trajectory obtained integrating $\mathbf{u}_1 + \mathbf{r}$ is the ``step''. By repeating this process it is possible to construct an arbitrary long pseudo-trajectory which follows the STCS. Using this trajectory we found that at $\epsilon = 0.199$, prior to the onset of permanent spatiotemporal chaos, the chaotic saddle STCS has a Lyapunov spectrum with 14 positive Lyapunov exponents and a Kaplan-Yorke dimension $\sim36$.

The edge state ES of Fig. \ref{fig_02}(b) is a saddle point in the Fourier space, with one positive Lyapunov eigenvalue and 39 negative Lyapunov eigenvalues, whose stable manifolds (EOC) separate the two regions of pseudo-basins in Fig. \ref{fig_01}(a) and accounts for the initial constant energy trajectory at the inset of Fig. \ref{fig_01}(b). The laminar structure of Fig. \ref{fig_02}(c), corresponding to a spatially regular and temporally chaotic attractor, has one positive Lyapunov exponent and a Kaplan-Yorke dimension of $\sim22$ \cite{chian2010}.

\begin{figure}[t]
\begin{center}
\includegraphics[width=1\linewidth]{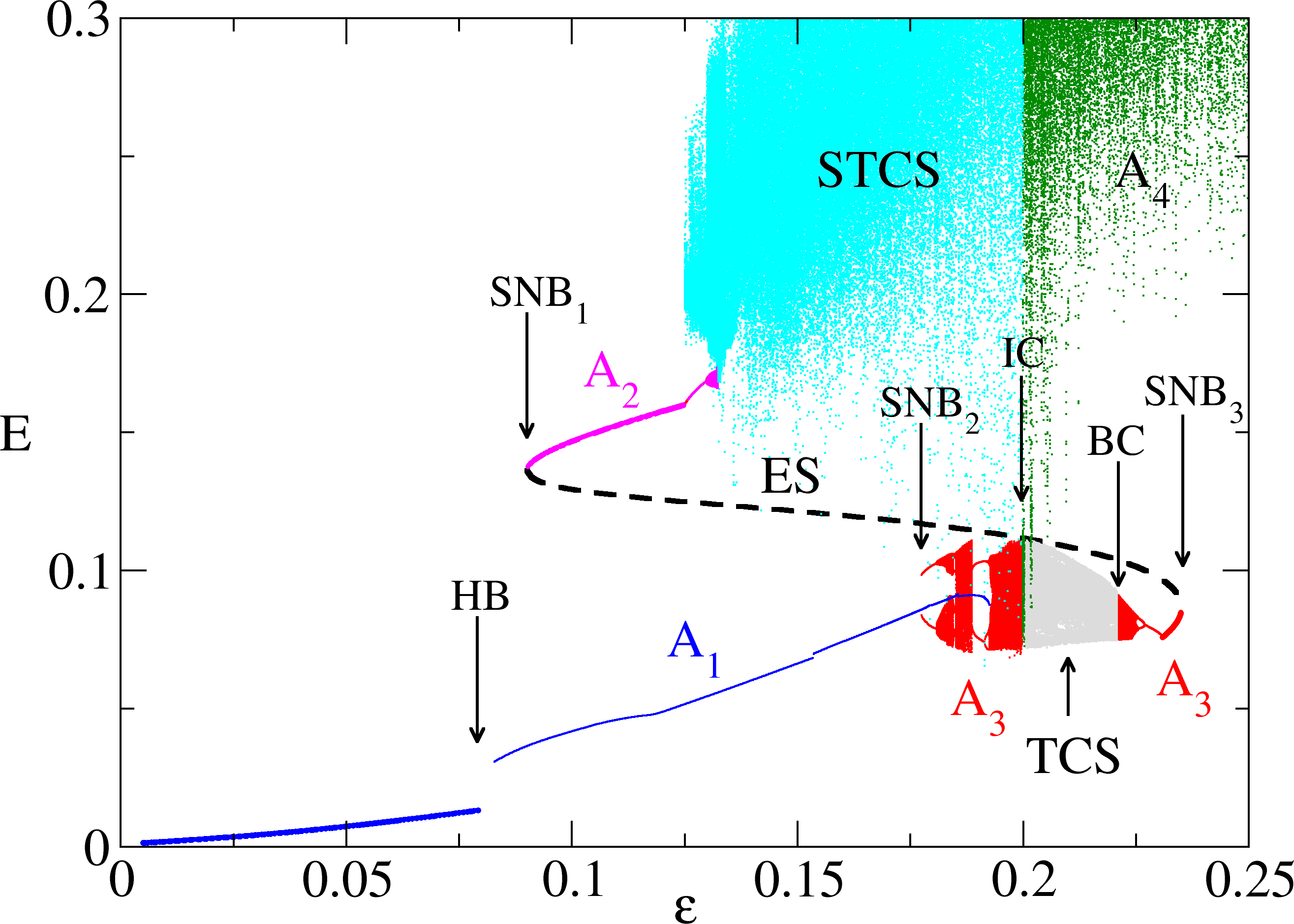} 
\end{center}
\caption{(Color online) Bifurcation diagram of $E$ as a function of $\epsilon$  for edge state (ES), attractors ($A_1$-$A_4$), spatiotemporally chaotic saddle (STCS) and temporally chaotic saddle (TCS), showing Hopf bifurcation (HB), saddle-node bifurcation (SNB), interior crisis (IC) and boundary crisis (BC).}
\label{fig_03}
\end{figure}
\subsection{The route to spatiotemporal chaos}
In order to probe the origin of the edge state and the genesis of transient turbulence related to the aforementioned EOC, we construct a detailed bifurcation diagram in Fig. \ref{fig_03} for $E$ as a function of $\epsilon$ for attractors, chaotic saddles and ES. We adopt a Poincar\'e map by plotting a point every time the trajectory obtained from Eq. \eqref{eq_01} crosses the plane $|\hat u_2(t)|=0.1$ with $d|\hat u_2(t)|/dt>0$. For $\epsilon=$0 to 0.25 we have identified four different attractors: $A_1$, $A_2$, $A_3$ and $A_4$. In the interval $\epsilon=0$ to 0.079 $A_1$ is a stable fixed point (thick blue line) with a constant energy for a given $\epsilon$,  which loses its stability and is converted to a period-1 limit cycle (thin blue line) via a Hopf bifurcation (HB) at $\epsilon\sim0.079$. We observe three small energy jumps in the $A_1$ branch, one of them is visible in Fig. \ref{fig_03} at $\epsilon\sim0.154$. These jumps represent transitions from one period-1 attractor to another. Although strictly speaking these are different attractors, we refer to them as $A_1$ in this paper because they occupy roughly the same area in the phase space and their bifurcations do not affect our main analysis. The last period-1 limit cycle vanishes at $\epsilon\sim 0.1925$. 

$A_2$ appears via a saddle-node bifurcation (SNB$_1$) at $\epsilon\sim0.09$ when two fixed points, one stable (thick magenta line) and one unstable (ES, dashed black line), are created. This unstable fixed point corresponds to the edge state that plays a fundamental role in the genesis of the EOC and transient turbulence seen in Figs. \ref{fig_01} and \ref{fig_03}, respectively. At $\epsilon\sim0.125$, $A_2$ suffers a Hopf bifurcation and becomes a limit cycle of period-1 (thin magenta line). At $\epsilon\sim0.1297$,  $A_2$ is bifurcated into a quasiperiodic attractor which loses its stability and vanishes at $\epsilon\sim0.13235$. Further research is required to determine the bifurcation that causes the disappearance of $A_2$.

The coexistence of attractors $A_1$ and $A_2$ in the interval $\epsilon\sim0.09$ to 0.13235 implies the existence of two basins of attraction. We will show that the dynamical changes of the basin boundary is responsible for the genesis of transient turbulence. 

A period-2 attractor $A_3$ (red line) appears via a saddle-node bifurcation (SNB$_2$) at $\epsilon\sim0.1774$, which undergoes a number of different bifurcations as $\epsilon$ increases, involving a transition to quasiperiodicity, period-doubling cascade, and unstable dimension variability to temporal chaos at $\epsilon\sim0.1925$ \cite{galuzio2010,*galuzio2011}, and at $\epsilon\sim0.2$ it loses its stability via an interior crisis (IC) leading to the onset of a spatiotemporally chaotic attractor $A_4$ (green) \cite{rempel2007}.

In references \cite{rempel2007,chian2010} it was demonstrated that for $\epsilon\lesssim0.21$ $A_4$ is composed of a spatiotemporally chaotic saddle which preexists as the transient turbulence prior to IC (STCS, light blue) and a temporally chaotic saddle (TCS, grey) evolved from $A_3$. TCS  turns into a temporally chaotic attractor $A_3$ at $\epsilon\sim0.22105$ due to a boundary crisis (BC). As $\epsilon$ increases further, $A_3$ turns into a period-1 limit cycle via an inverse period-doubling cascade and becomes a stable fixed point (thick red line) via a Hopf bifurcation at $\epsilon\sim0.2308$. At $\epsilon\sim 0.235$,  the stable fixed point $A_3$ disappears in a saddle-node bifurcation (SNB$_3$), along with ES. $A_3$ and $A_4$ coexist for $\epsilon\sim0.22105$ to 0.235. 

\begin{figure}[t]
\begin{center}
\includegraphics[width=1\linewidth]{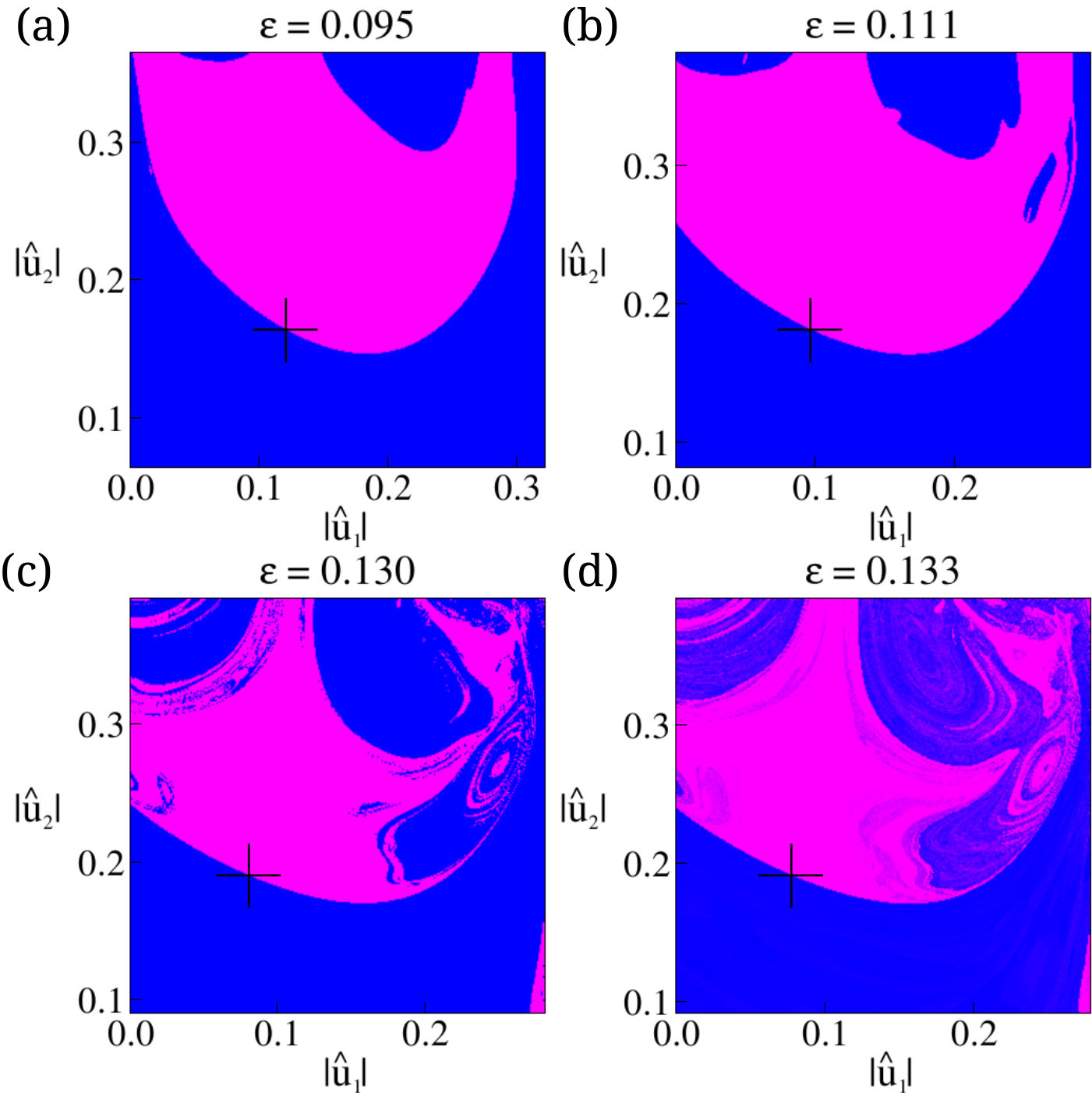} 
\end{center}
\caption{(Color online) Basins of attraction for the coexisting attractors $A_1$ (blue) and $A_2$ (magenta) at $\epsilon = 0.095$ (a), 0.111 (b), 0.130 (c). The black cross denotes the edge state. (d) Thresholded lifetime function of $A_1$ at $\epsilon = 0.133$, just after the disappearance of
$A_2$ . Blue (magenta) regions indicate initial conditions with lifetime shorter (longer) than the mean lifetime.}
\label{fig_04}
\end{figure}

We detect the STCS for $0.125\lesssim\epsilon\lesssim0.2$ using the sprinkler method \cite{kantz1985,*hsu1988}. Detection of chaotic saddles using the sprinkler method is based upon the ability to find initial conditions with long chaotic transients, but we found out that for $\epsilon<0.125$ the transient times become too short to be useful for detecting the chaotic saddles. However, as we argue below, the STCS is present from $\epsilon\approx0.11$. Therefore, EOC can be found for $0.11\lesssim \epsilon \lesssim 0.2$ where there is coexistence of transient turbulence (STCS), edge state (ES) and spatially regular laminar attractors ($A_1$, $A_2$, $A_3$).

Next we investigate the origin of the edge state and its role in the genesis and evolution of STCS responsible for the transient turbulence. As mentioned earlier, when $A_2$ appears as a stable fixed point, an unstable fixed point appears simultaneously via SNB$_1$; $A_1$ coexists with $A_2$ for $0.09\lesssim\epsilon\lesssim0.13235$. The basins of attraction are separated by a boundary. ES is a saddle structure that lies at the basin boundary. We applied the bisection method \cite{skufca2006} to detect ES that separates  $A_1$ and $A_2$, and discovered that ES is the unstable fixed point born at the saddle-node bifurcation SNB$_1$ ($\epsilon\sim0.09$), shown in Fig. \ref{fig_03}. The edge state corresponds to an unstable travelling wave moving with the driver speed in the laboratory frame.

\begin{figure}[t]
\begin{center}
\includegraphics[width=0.8\linewidth]{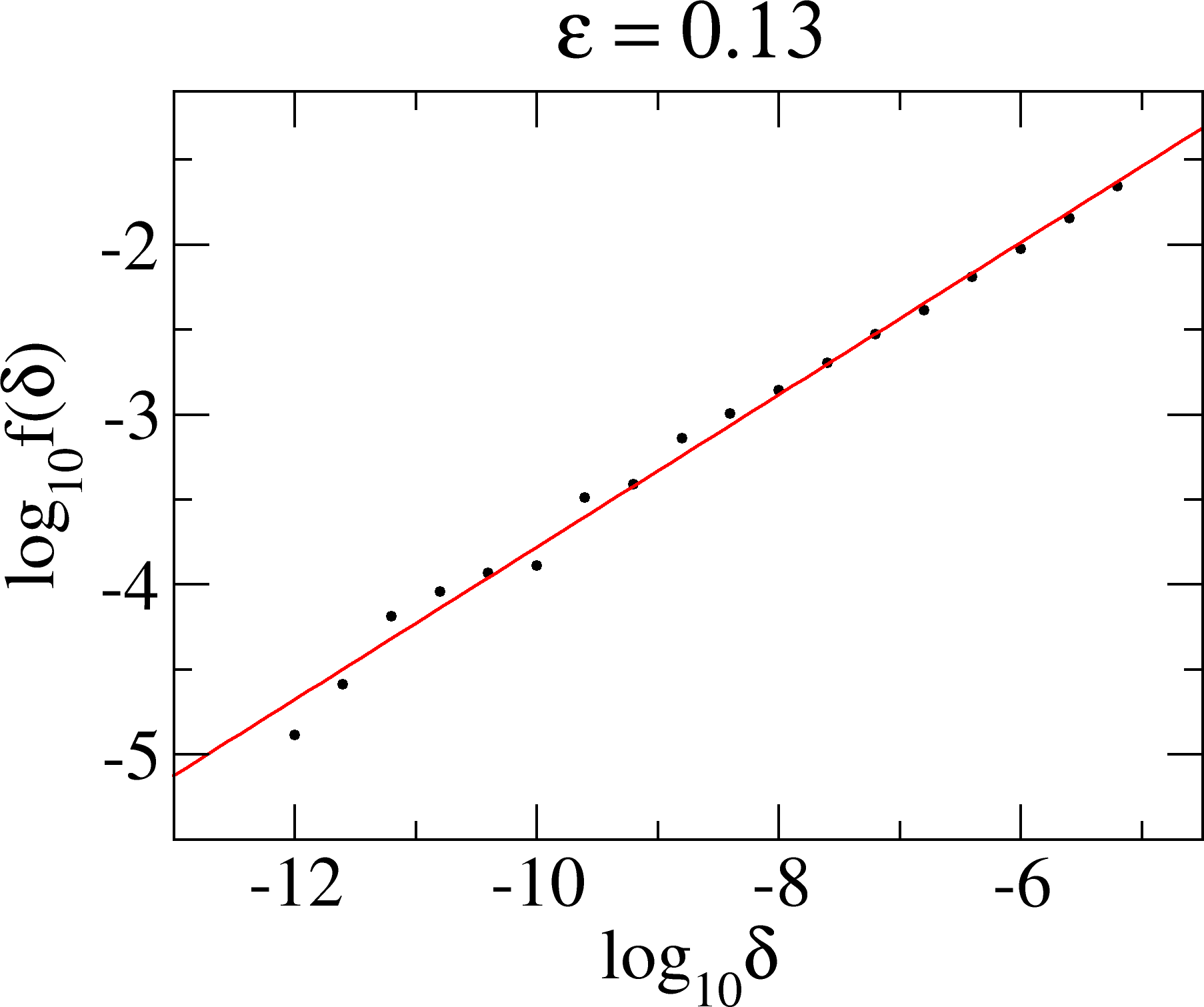} 
\end{center}
\caption{(Color online) Log-log plot of the uncertain fraction $f$ versus the error $\delta$ at $\epsilon=0.13$ (black dots). The red straight line fits the data set, in agreement with Eq. \eqref{eq_05}.}
\label{fig_05}
\end{figure}
\subsection{Genesis of edge state and transient turbulence}
Figure \ref{fig_04} shows the basins of attraction of $A_1$ (blue) and $A_2$ (magenta) for $\epsilon= 0.095$, 0.111, and 0.130, respectively, along with the edge state (cross). The stable manifold of ES is the basin boundary separating 2 coexisting attractors. For  $\epsilon  < \epsilon_\text{M}\sim0.11$, the basin boundary is smooth as shown by Fig. \ref{fig_04}(a) for $\epsilon = 0.095$. At $\epsilon =  \epsilon_\text{M}$, the basin boundary suffers a smooth-fractal metamorphosis \cite{grebogi1986, *grebogi1987, *tel_lai2011}, as seen in Fig. \ref{fig_04}(b), and a chaotic saddle STCS is born in this process. As $\epsilon$ increases for $\epsilon  > \epsilon_\text{M}$ the basin boundaries become increasingly complex due to a cascade of fractal-fractal metamorphoses, as shown in Fig. \ref{fig_04}(c).
\subsection{Characterization of regimes}
\subsubsection{The uncertainty exponent}
\begin{figure}[t]
\begin{center}
\includegraphics[width=1\linewidth]{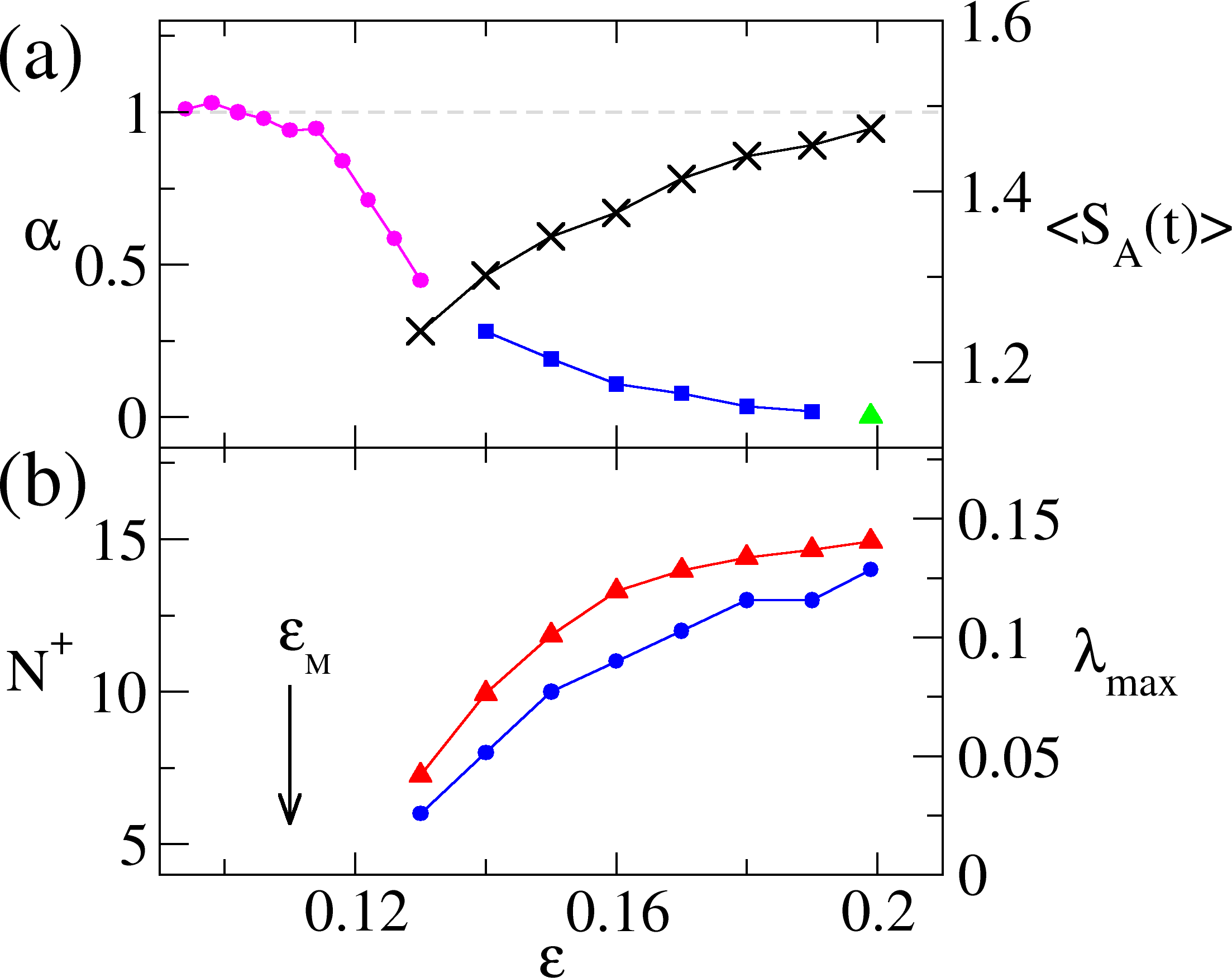} 
\end{center}
\caption{(Color online) Variation with $\epsilon$ of: a) the uncertainty exponent $\alpha$ (magenta circles, blue squares, green triangle) and the time-average power spectral entropy $\langle S_A(t)\rangle$ (cross), b) the number of positive Lyapunov exponents $N^+$ (blue circles) and the maximum Lyapunov exponent $\lambda_\text{max}$ (red triangles). $\epsilon_M$ indicates the genesis of a chaotic saddle (STCS).}
\label{fig_06}
\end{figure}

The uncertainty exponent is a mathematical method of measuring the fractal dimension of a basin boundary. Since the transient turbulence is related to a chaotic saddle located at a basin boundary, which is non-attracting and of measure zero, a practical way to infer and measure the properties of STCS is through the basin boundaries \cite{grebogi1983}. To measure the fractal dimension of the basin boundary for $0.09\lesssim\epsilon\lesssim0.13235$, where $A_1$ and $A_2$ coexist, we compute the uncertainty exponent $\alpha = D-d$, where $D$ is the dimension of the phase space and $d$ is the fractal dimension of the basin boundary, with $\alpha$ defined between 0 (total fractality) and 1 (smooth). Here, by total fractality we mean that the dimension of the fractal boundary approaches the dimension of the phase space. The uncertainty exponent is related to the uncertainty fraction \cite{grebogi1983,*moon1985} 
\begin{equation}
  f(\delta) \sim \delta^\alpha, 
\label{eq_05}
\end{equation}
where $f(\delta)$ is the fraction of uncertain initial conditions with respect to the error $\delta$. In the case of coexistence of attractors, an initial condition $\mathbf{u}$ is classified as uncertain with respect to $\delta$ if the perturbed initial condition $\mathbf{u}+\delta\hat{\mathbf{e}}$ converges to another attractor different from $\mathbf{u}$, with $\hat{\mathbf{e}}$ being an arbitrary unit vector. To ensure good statistical convergence, we classify initial conditions randomly until reaching a number of 500 uncertain conditions. Then, we divide the number of uncertain conditions by the total number of initial conditions chosen to obtain the uncertain fraction. We repeat this procedure for different values of $\delta\in[10^{-12},10^{-5}]$ and determine $\alpha$ from the slope of the straight line which fits $f(\delta)$ vs $\delta$ in a log-log scale. As an example, black dots in Fig. \ref{fig_05} are the uncertain fraction $f(\delta)$ for $\epsilon=0.13$. The red straight line is obtained from the linear regression analysis of the data set, with a slope $\alpha=0.45\pm0.01$. The computed $\alpha$ for the interval where attractors $A_1$ and $A_2$ coexist is shown by the magenta circles in Fig. \ref{fig_06}(a). 

\begin{figure}[t]
\begin{center}
\includegraphics[width=1\linewidth]{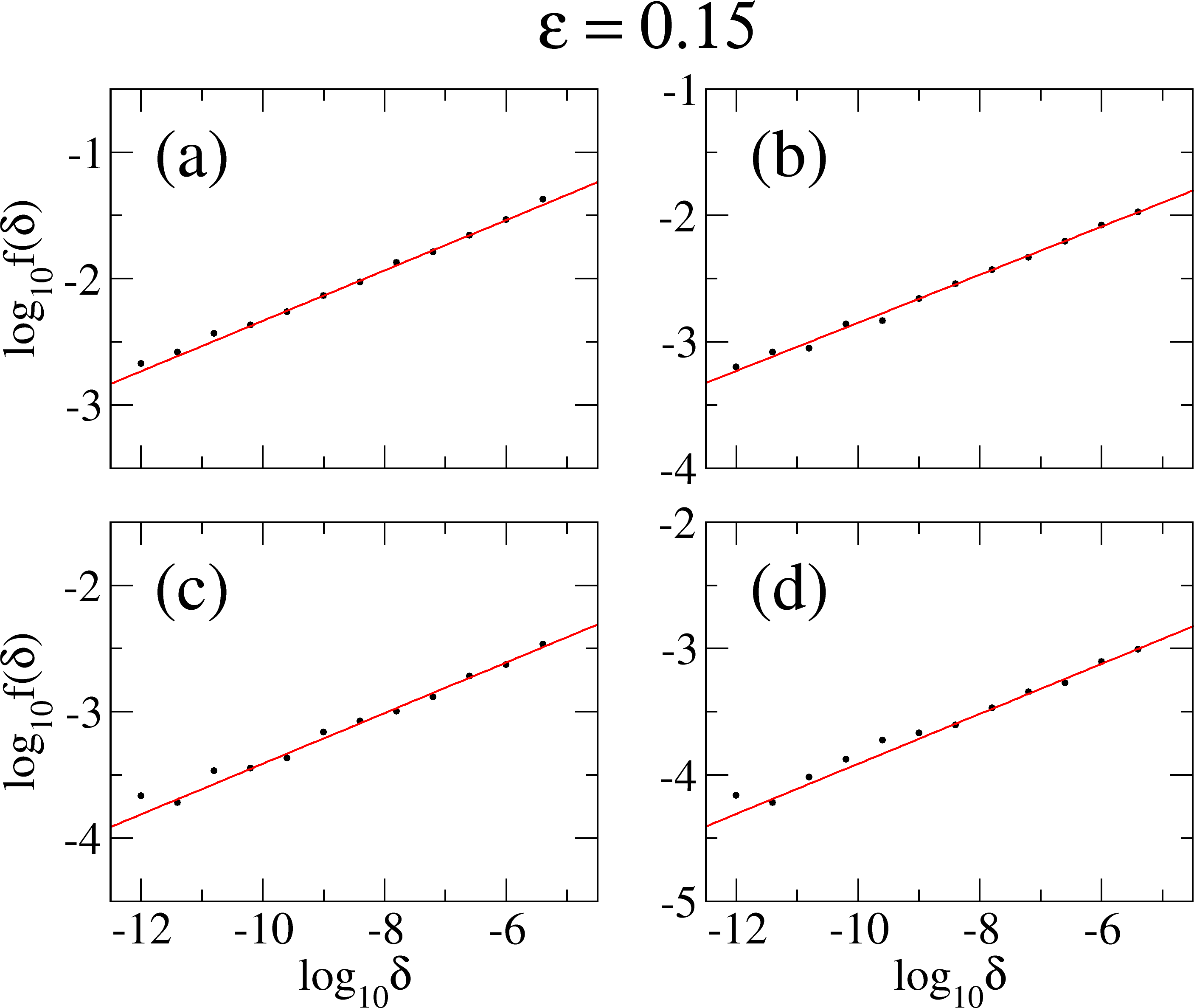} 
\end{center}
\caption{(Color online) Log-log plots of the uncertain fraction $f$ versus the error $\delta$ at $\epsilon=0.15$, computed using four different time-difference thresholds: (a) $\Delta T^*$=10, (b) $\Delta T^*$=20, (c) $\Delta T^*$=30 and (d) $\Delta T^*$=40. Time is in Poincar\'e map units.}
\label{fig_07}
\end{figure}

Figure \ref{fig_04}(d) shows the thresholded lifetime function of $A_1$ at $\epsilon = 0.133$, just after the disappearance of $A_2$ , in the same region of the phase space of Fig. \ref{fig_04}(c). Blue (magenta) regions indicate initial conditions with lifetime shorter (longer) than the mean lifetime. We use this information to compute the uncertainty exponent of the pseudo-basin boundary for $0.14\leq \epsilon\leq0.19$. Based on the method first introduced by \citet{lau1991}, we define the lifetime difference for 2 initial conditions separated by a distance $\delta$ as 
\begin{equation}
\Delta T(\mathbf{u}) = |T(\mathbf{u} + \delta\hat{\mathbf{e}})- T(\mathbf{u})|, 
\label{eq_06}
\end{equation}
where $T(\mathbf{u})$ is the time an initial condition $\mathbf{u}$ takes to converge to $A_1$. We classify an initial condition $\mathbf{u}$ as uncertain if $\Delta T(\mathbf{u}) > \Delta T^*$, where $\Delta T^*$ is a positive time-difference threshold. Following the same procedure decribed above, we obtain $f(\delta)$ for many values of $\delta$. According to \citet{aguirre2009}, the uncertainty exponent obtained in this way does not depend on the value of $\Delta T^*$, as long as the threshold is not too small. Figure \ref{fig_07} shows the log-log plot of $f(\delta)$ versus $\delta$ for $\Delta T^*=10, 20, 30$ and 40, in units of Poincar\'e map iterations, at $\epsilon=0.15$. Applying a linear regression analysis, we found that all data sets in Fig. \ref{fig_07} are well fitted by straight lines, as quantified by the linear correlation coefficients $r^2$ close to 1, in agreement with Eq. \eqref{eq_05}. Furthermore, the uncertainty exponent $\alpha$ is very similar for all $\Delta T^*$, with low standard deviations $\Delta\alpha$. The information obtained from linear regressions is summarized in Table \ref{tab_01}. To compute $\alpha$ in the interval $0.14\leq \epsilon\leq0.19$, we produce data sets similar to those of Fig. \ref{fig_07} for different values of $\epsilon$ in the interval and keep the value of $\alpha$ associated with the data set which better fits a straight line, given by the linear correlation coefficient $r^2$ closer to 1. The values of $\alpha$ calculated for the interval $0.14\leq \epsilon\leq0.19$ are indicated by the blue squares in Fig. \ref{fig_06}(a).

\begingroup
\begin{table}[b]
\caption{Information from linear regressions applied to the data sets in Fig. \ref{fig_07}.}
\label{tab_01}
\begin{tabular}{|c|c|c|}\hline
$\Delta T^*$ & $\alpha\pm\Delta\alpha$ & $r^2$ \\ \hline
10 &$0.194\pm0.004$&0.997 \\
20 &$0.189\pm0.004$&0.995 \\
30 &$0.188\pm0.007$&0.985 \\
40 &$0.184\pm0.007$&0.987 \\ \hline
\end{tabular}
\end{table}
\endgroup

To compute the fractal dimension of the pseudo-basin boundary for $\epsilon=0.199$, we use the expression for the upper bound of the uncertainty exponent as a function of the mean lifetime of the chaotic saddle $\tau$ \cite{rempel2007,tel_lai2011} and the maximum Lyapunov exponent $\lambda_\text{max}$  \cite{tel1990,{*paar2000}},
\begin{equation}
  \alpha \leq \frac{1}{\tau\lambda_\text{max}}.
\label{eq_07}
\end{equation}
\citet{chian2010} obtained $\lambda_\text{max}$ = 0.1405 and $\tau = 353.3$ (in units of the driver period) for $\epsilon = 0.199$, thus $\alpha$ calculated from the upper bound formula is 0.002018, denoted by the green triangle in Fig. \ref{fig_06}(a).

Figure \ref{fig_06}(a) shows that for $\epsilon\lesssim 0.11$, $\alpha$ is approximately 1, implying a smooth basin boundary. For $\epsilon\gtrsim 0.11$, $\alpha$ becomes less than 1, implying a fractal basin boundary. As $\epsilon$ increases further, $\alpha$ steadily decreases, implying the occurrence of a cascade of fractal-fractal metamorphoses \cite{grebogi1986}, in agreement with changes in the basin boundaries seen in Fig. \ref{fig_04}. After the disappearance of attractor $A_2$, the fractal dimension of the pseudo-basin boundary increases continuously as $\epsilon$ increases, as shown by the blue squares in Fig. \ref{fig_06}(a). At $\epsilon=0.199$, the fractal dimension of the pseudo-basin boundary, corresponding to the dimension of the stable manifold of the STCS, reaches a value near the dimension of the phase space ($\alpha\sim0$). This can be seen as the intermingled blue region in Fig. \ref{fig_01}(a). As noted by \citet{lai1995}, this feature of the pseudo-basin is related to long spatiotemporal chaotic transients in spatially-extended systems due to the presence of a chaotic saddle.
\subsubsection{The degree of complexity}
To complement the characterization of regimes based on the variation of the uncertainty exponent $\alpha$ as a function of the control parameter $\epsilon$, we compute three different quantifiers of the degree of complexity: (i) the time-average of the Fourier power spectral Shannon entropy $\langle S_A(t)\rangle$, (ii) the number of positive Lyapunov exponents $N^+$, and (iii) the maximum Lyapunov exponent $\lambda_{\text{max}}$. The Fourier power spectral Shannon entropy quantifies the degree of amplitude synchronization between Fourier modes \cite{rempel2007}. The number of positive Lyapunov exponents, obtained from the Lyapunov spectrum, can be used to measure the degree of spatiotemporal chaos or turbulence. For example, the laminar attractor has only one positive Lyapunov exponent whereas the chaotic saddle associated with the transient turbulence may have up to 14 positive Lyapunov exponents \cite{chian2010}, which is consistent with degrees of amplitude synchronization quantified by $\langle S_A(t)\rangle$. The maximum Lyapunov exponent quantifies the degree of temporal chaoticity of the system.

In order to characterize the spatiotemporal dynamics of the chaotic saddle STCS created at $\epsilon\gtrsim0.11$ as a function of $\epsilon$, first we generate arbitrarily long trajectories near the STCS by using the stagger-and-step method \cite{sweet2001}. To quantify the degree of spatial disorder of the STCS we compute the time-average of the Fourier power spectral Shannon entropy \cite{chian2010}, given by 
\begin{equation}
S_A(t)=-\sum_{k=1}^N p_k(t)\ln p_k(t), 
\label{eq_08}
\end{equation}
where $p_k(t)$ is the relative wight of a Fourier mode $k$ at an instant $t$
\begin{equation}
p_k(t)=|\hat{u}_k(t)|^2/\sum_{k=1}^N|\hat{u}_k(t)|^2.
\label{eq_09}
\end{equation}
Figure \ref{fig_06}(a) shows that the degree of spatial disorder increases with $\epsilon$ until it reaches the maximum value near $\epsilon\sim0.199$. 

Moreover, we compute the Lyapunov spectrum \cite{shimada1979, *benettin1980} solving the variational equation for the flux Jacobian matrix from the STCS trajectories (see \citet{miranda2012} for further details). The increase of the degree of spatial disorder with increasing driver amplitude is accompanied by an increase of temporal chaos. Figure \ref{fig_06}(b) shows that the number of positive Lyapunov exponents $N^+$ (blue circles) increases steadily with increasing $\epsilon$, reaching its maximum value of $N^+ = 14$ at $\epsilon \sim 0.199$. Similar behavior is observed for the maximum Lyapunov exponent $\lambda_\text{max}$ (red triangle), shown in Fig. \ref{fig_06}(b). Figure \ref{fig_06} provides a consistent overview of the genesis and evolution of the transient turbulence showing that the degree of complexity of transient turbulence (STCS) increases as $\epsilon$ increases and evolves to a well-developed transient turbulence before the transition to permanent spatiotemporal chaos.

\section{Conclusion}

We have demonstrated that prior to the onset of permanent spatiotemporal chaos the regularized long-wave equation exhibits the behavior of edge of chaos, whereby a trajectory traverses a transient turbulent state before converging to a laminar state. The edge state responsible for the EOC and the genesis of turbulence was identified and a sequence of metamorphoses of the EOC was shown to be responsible for the appearance of a chaotic saddle and its subsequent evolution to a well-developed transient turbulence. Our results provide a much clearer picture of the origin of turbulence in the regularized long-wave equation, which has been extensively studied as a general model of transition to spatiotemporal chaos \cite{he2003,rempel2007,he1988,galuzio2010}. These results can be applied to a wide class of spatially extended systems where a transient turbulence (STCS) coexists with laminar (spatially regular) attractors before transition to an asymptotic turbulence \cite{rempel2007b,*rempel2009,*rempel2010}.

\begin{acknowledgements}
This paper is dedicated to Judith Ling, MD (Stanford), for her strong devotion to the needy and underprivileged, and for her contribution to tropical disease research. This work is supported by CAPES, CNPq and FAPESP. A.C.L.C. thanks European Commission for the award of a Marie Curie International Incoming Fellowship and the hospitality of Paris Observatory. P.R.M. thanks FAPESP, Process 2011/10466-1, for a post-doctoral fellowship. The authors thank O. M. Podvigina and V. A. Zheligovsky for valuable discussions, and five referees for comments.
\end{acknowledgements}

\bibliography{references}

\end{document}